\documentstyle[preprint,aps]{revtex}

\input psfig.sty
\tightenlines 

\begin{document}
\draft
\title{Direct Urca processes in superdense cores of neutron stars\thanks{%
Talk given at the International Workshop ``HADRON PHYSICS 2002'', April 14
-- 19, Bento Gon\c{c}alves, Rio Grande do Sul, Brazil.}}
\author{L. B. Leinson}
\address{Institute of Terrestrial Magnetism, Ionosphere and Radio Wave\\
Propagation RAS, 142190 Troitsk, Moscow Region, Russia\\
E-mail: leinson@izmiran.rssi.ru}
\maketitle

\begin{abstract}
We use the field theoretical model to perform relativistic calculations of
neutrino energy losses caused by the direct Urca processes on nucleons in
the degenerate baryon matter. By our analysis, in a free nucleon gas under
beta equilibrium, the direct neutron decay is forbidden if the number
density of neutrons exceeds the critical value $n_{n}^{c}=5.9\times
10^{31}\,cm^{-3}$. In superdense nuclear matter, $n>n_{0}$, the weak decay
of neutrons is possible only due to strong interactions, caused by an
exchange of isovector mesons. Mean field of isovector mesons in the medium
creates a large energy gap between spectrums of protons and neutrons, which
is required by kinematics of beta decay. Our expression for the neutrino
energy losses, obtained in the mean field approximation, incorporates the
effects of nucleon recoil, parity violation, weak magnetism, and
pseudoscalar interaction. For numerical testing of our formula, we use a
self-consistent relativistic model of the multicomponent baryon matter. The
relativistic emissivity of the direct Urca reactions is found substantially
larger than predicted in the non-relativistic approach. We found that, due
to weak magnetism effects, relativistic emissivities increase by
approximately 40-50\%.
\end{abstract}


\widetext

The simplest neutrino-emitting processes by which neutron stars can lose
their thermal energy are beta decay of neutrons and electron capture on
protons $n\rightarrow p+l+\bar{\nu}_{l},\ \ \ p+l\rightarrow n+\nu _{l}$.
These reactions, widely known as the direct Urca processes, are a central
point of any modern scenarios of evolution of neutron stars because the
neutrino energy losses caused by the direct Urca processes lead to a rapid
cooling of the degenerate neutron star core. In the non-relativistic
approximation, the energy losses caused by the direct Urca processes have
been calculated more than ten years ago\cite{Lat91}%
\begin{equation}
Q_{nr}=\frac{457\pi }{10080}G_{F}^{2}C^{2}\left( C_{V}^{2}+3C_{A}^{2}\right)
T^{6}M_{n}^{\ast }M_{p}^{\ast }\mu _{e}\Theta \left( p_{e}+p_{{\rm p}}-p_{%
{\rm n}}\right) ,  \label{QLnr}
\end{equation}%
Here $G_{F}$ is the Fermi weak coupling constant; $C=\cos \theta
_{C}=0\allowbreak .\,\allowbreak 973$ is the Cabibbo factor; $C_{V}=1$ and $%
C_{A}\simeq 1.26$ are the nucleon weak coupling constants. Here $M_{n}^{\ast
}$ and $M_{p}^{\ast }$ are the effective nucleon masses, and $\mu _{e}$ is
the electron chemical potential.

The energy exchange in the matter goes naturally on the temperature scale $%
\sim T$, which is small compared to typical kinetic energies of degenerate
particles. Therefore the momenta of in-medium fermions are fixed at their
values at Fermi surfaces, which we denote as $p_{{\rm n}}$, $p_{{\rm p}}$
for the nucleons and $p_{e}$ for electrons respectively. Then the
''triangle'' condition, $p_{{\rm p}}+p_{e}>p_{{\rm n}}$, required by the
step-function, is necessary for conservation of the total momentum in the
reaction and, together with the condition of charge neutrality of the
medium, results in the threshold dependence on the proton concentration.

In spite of widely adopted importance of the direct Urca reactions, the
corresponding neutrino energy losses are not well investigated yet. The
simple formula (\ref{QLnr}) has been derived in the non-relativistic manner
by neglecting all interactions between nucleons. Actually, however, the
superthreshold proton fraction, necessary for the direct Urca processes to
operate in the degenerate nuclear matter, appears at large densities, where
the Fermi momenta of participating nucleons are comparable with their
effective mass. Moreover, according to modern numerical simulations, the
central density of the star can be up to eight times larger than the nuclear
saturation density. This implies a substantially relativistic motion of
nucleons in the superdense neutron star core. The appropriate equation of
state for such a matter is actually derived in the relativistic approach,
and the relevant neutrino energy losses must be consistent with the used
relativistic equation of state.

We represent the totally relativistic consideration of the direct Urca
processes on nucleons. Let us consider first the relativistic kinematics of
the neutron beta decay $n\rightarrow p+l+\bar{\nu}_{l}$ in the degenerate
matter under beta equilibrium. In what follows we consider massless
neutrinos of energy and momentum $k_{1}=\left( \omega _{1},{\bf k}%
_{1}\right) $ with $\omega _{1}=\left| {\bf k}_{1}\right| .$ The
energy-momentum of the final lepton $l=e^{-},\mu ^{-}$ of mass $m_{l}$ is
denoted as $k_{2}=\left( \omega _{2},{\bf k}_{2}\right) $ with $\omega _{2}=%
\sqrt{{\bf k}_{2}^{2}+m_{l}^{2}}$. Thus, the energy and momentum
conservation in the beta decay is given by the following equations%
\begin{eqnarray}
E_{{\rm n}}\left( {\bf p}\right) -E_{{\rm p}}\left( {\bf p}^{\prime }\right)
-\omega _{1}-\omega _{2} &=&0  \nonumber \\
{\bf p}-{\bf p}^{\prime }-{\bf k}_{1}-{\bf k}_{2} &=&0,  \label{cons}
\end{eqnarray}%
where $E_{{\rm n}}\left( {\bf p}\right) $ and $E_{{\rm p}}\left( {\bf p}%
^{\prime }\right) $ are the in-medium energies of the neutron and the proton
respectively.

Since the antineutrino energy is $\omega _{1}\sim T$, and the antineutrino
momentum $\left| {\bf k}_{1}\right| \sim T$ is much smaller than the momenta
of other particles, we can neglect the neutrino contributions in Eqs. (\ref%
{cons}). Then the momentum conservation, ${\bf p}_{{\rm p}}+{\bf p}_{l}={\bf %
p}_{{\rm n}}$, implies the well known ''triangle'' condition, $p_{{\rm p}%
}+p_{l}>p_{{\rm n}},$necessary for the Urca processes to operate. However,
the momentum conservation is necessary but insufficient condition for the
direct beta decay to occur. The energy conservation requires a one more
condition%
\begin{equation}
E_{{\rm n}}\left( p_{{\rm n}}\right) -E_{{\rm p}}\left( p_{{\rm p}}\right) =%
\sqrt{\left( {\bf p}_{{\rm n}}-{\bf p}_{{\rm p}}\right) ^{2}+m_{l}^{2}},
\label{Econs}
\end{equation}%
Squaring of both sides of this equation gives%
\begin{equation}
\left( E_{{\rm n}}\left( p_{{\rm n}}\right) -E_{{\rm p}}\left( p_{{\rm p}%
}\right) \right) ^{2}-\left( {\bf p}_{{\rm n}}-{\bf p}_{{\rm p}}\right)
^{2}=m_{l}^{2}.  \label{Ec}
\end{equation}%
If we denote as 
\begin{equation}
K=\left( E_{{\rm n}}-E_{{\rm p}},{\bf p}_{{\rm n}}-{\bf p}_{{\rm p}}\right) ,
\label{K}
\end{equation}%
the energy-momentum transferred from the nucleon, this equation can be
readily recognized as the following condition $K_{\mu }^{2}=m_{l}^{2}$. This
condition naturally arises from the time-like momentum of the final lepton
pair, $K=k_{1}+k_{2}\simeq k_{2}$, and can be satisfied only in the presence
of the energy gap between spectrums of protons and neutrons. Really,
consider the neutron decay in the degenerate matter. In this case the
neutron from the neutron Fermi surface $E_{Fn}$ undergoes transition into
the proton at the proton Fermi energy $E_{Fp}$. \vskip0.3cm %
\psfig{file=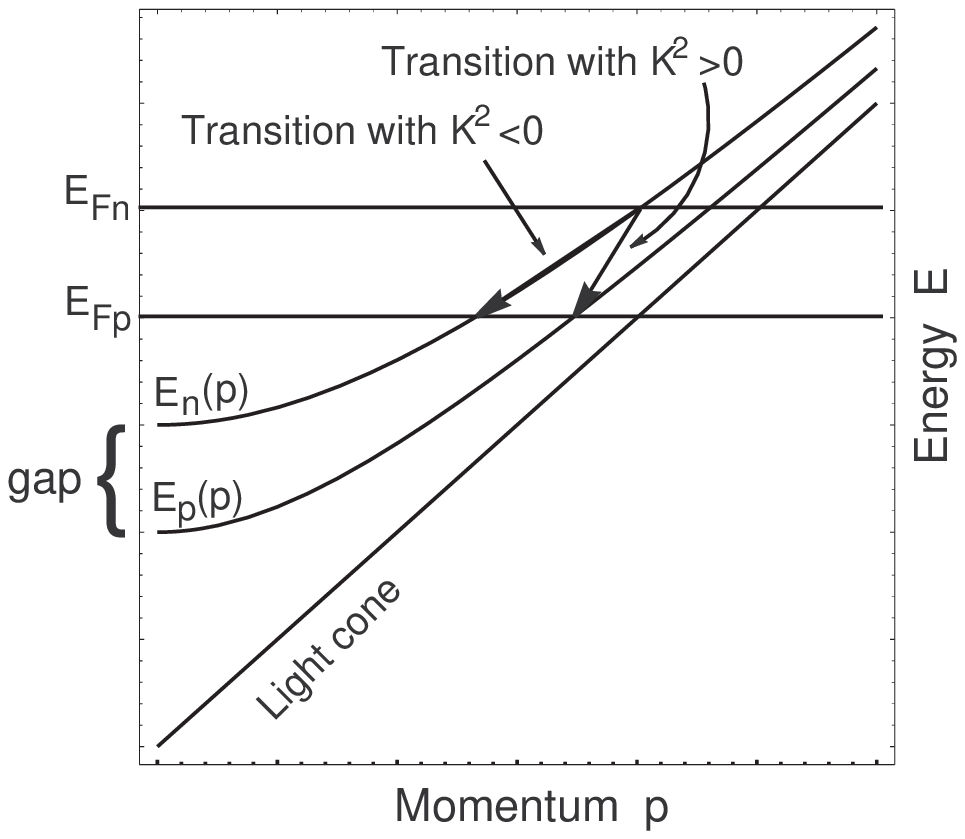}Fig. 1. The nucleon transition caused bu the beta decay.
If the neutron and proton energy spectrums are identical (the upper curve),
the momentum transfer due to such a transition is space-like. The energy gap
between protons and neutrons allows the time-like momentum transfer to the
leptons. \vskip0.3cm If the neutron and proton energy spectrums are
identical (the upper curve, for example), the momentum transfer due to such
a transition is space-like contrary to the time-like momentum of the final
lepton pair. In this case the neutron decay is kinematically forbidden. In
contrast, the energy gap between protons and neutrons allows the time-like
momentum transfer to the leptons, and thus opens the direct neutron decay.
It should be emphasized however that the stated conditions are consistent
only in the presence of the energy gap between protons and neutrons.
Therefore the possibility of the direct neutron decay depends substantially
on the model of nuclear matter.

Consider, for example, a degenerate free gas consisting of neutrons,
protons, and electrons under beta equilibrium. In this case the energy gap
exists only due to the mass difference of neutron and proton. If we denote
the masses as $M_{{\rm n}}$ and $M_{{\rm p}}$ respectively, then the
corresponding Fermi energies are of the following relativistic form 
\[
E_{{\rm n}}\left( p_{{\rm n}}\right) =\sqrt{M_{{\rm n}}^{2}+p_{{\rm n}}^{2}}%
,\ \ \ \ E_{{\rm p}}\left( p_{{\rm p}}\right) =\sqrt{M_{{\rm p}}^{2}+p_{{\rm %
p}}^{2}}. 
\]%
Due to charge neutrality the number density of electrons, $n_{e}\propto
p_{l}^{3}$, equals to the number density of protons, $n_{{\rm p}}\propto p_{%
{\rm p}}^{3}$. This implies equality of the proton and electron Fermi
momenta, $p_{{\rm e}}=p_{{\rm p}}$, so that the ''triangle'' condition reads 
$2p_{{\rm p}}=p_{{\rm n}}$. Neutrinos and antineutrinos freely escape from
the neutron star. Then the chemical potential of neutrinos equals zero, $\mu
_{\nu }=0$, and the condition of chemical equilibrium can be written as $\mu
_{p}+\mu _{e}=\mu _{n}$. Chemical potentials of degenerate particles can be
approximated by their individual Fermi energies, yielding 
\begin{equation}
\sqrt{m_{{\rm e}}^{2}+p_{{\rm p}}^{2}}+\sqrt{M_{{\rm p}}^{2}+p_{{\rm p}}^{2}}%
=\sqrt{M_{{\rm n}}^{2}+p_{{\rm n}}^{2}}.  \label{consF}
\end{equation}%
Solution of this equation 
\begin{equation}
p_{{\rm p}}\left( p_{{\rm n}}\right) =\frac{1}{2\sqrt{M_{{\rm n}}^{2}+p_{%
{\rm n}}^{2}}}\sqrt{\left( p_{{\rm n}}^{2}+M_{{\rm n}}^{2}-\left( M_{{\rm p}%
}+m_{{\rm e}}\right) ^{2}\right) \left( p_{{\rm n}}^{2}+M_{{\rm n}%
}^{2}-\left( M_{{\rm p}}-m_{{\rm e}}\right) ^{2}\right) \allowbreak }
\label{pbeta}
\end{equation}%
$\allowbreak $ gives the proton Fermi momentum as a function of neutron
Fermi momentum in the beta-equilibrated gas of protons and neutrons. This
function is shown by solid line. \vskip0.3cm \psfig{file=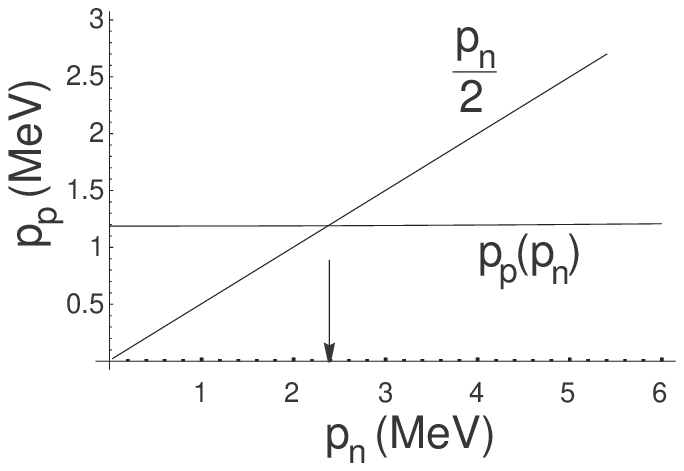}Fig. 2.
Solid line is the proton Fermi momentum as a function of neutron Fermi
momentum in the beta-equilibrated gas of protons and neutrons. Long-dashed
line is $p_{{\rm n}}/2$. The arrow shows the critical value of neutron Fermi
momentum above which the direct neutron decay is forbidden. \vskip0.3cm As
required by the ''triangle'' condition, $2p_{{\rm p}}\geq p_{{\rm n}}$, the
direct neutron decay in such a medium is open when the proton Fermi
momentum, $p_{{\rm p}}$, is larger than one half of the neutron Fermi
momentum, $p_{{\rm n}}/2$. This function is shown by long-dashed line. We
can see, the direct neutron decay is allowed to the left of the crossing
point and is forbidden at the right-hand side. The crossing point can be
found as solution of the following equation%
\begin{equation}
p_{{\rm p}}\left( p_{{\rm n}}\right) =\frac{p_{{\rm n}}}{2}  \label{1}
\end{equation}%
$\allowbreak $That is the critical value of the neutron Fermi momentum above
which the direct neutron decay is forbidden 
\begin{equation}
p_{{\rm n}}^{c}=\frac{\sqrt{\left( \left( M_{p}+M_{n}\right)
^{2}-m_{l}^{2}\right) \left( \left( M_{n}-M_{p}\right) ^{2}-m_{l}^{2}\right) 
}}{\sqrt{2M_{p}^{2}+2m_{l}^{2}-M_{n}^{2}}}=2.\,\allowbreak 381\,1\,MeV.
\label{2}
\end{equation}%
Thus the direct neutron decay is forbidden if the number density of neutrons
exceeds some critical value shown here 
\begin{equation}
n_{n}^{c}=\frac{\left( p_{{\rm n}}^{c}\right) ^{3}}{3\pi ^{2}}=5.932\times
10^{31}\,cm^{-3}  \label{3}
\end{equation}%
This number density is much smaller than that typical for neutron star
cores, therefore the direct neutron decay in the cooling neutron stars can
occur only due to strong interactions.

Notice, some of models of strong interaction also forbid the direct neutron
decay. Consider, for example, a simple model for the baryon matter, which
contains fields for baryons and neutral scalar $\left( \sigma \right) $ and
vector $\left( \omega _{\mu }\right) $ mesons. The isoscalar mesons equally
interact with protons and neutrons, which therefore have identical energy
spectrums. As discussed, in this case, the energy and the momentum in the
direct neutron decay can not be conserved simultaneously.

To allow the direct neutron decay, the model of nuclear matter must be
generalized to include some additional degrees of freedom and couplings,
which are able to create a large energy gap between possible energies of
protons and neutrons. For this purpose, besides isoscalar mesons $\sigma $
and $\omega $, the model should include also isovector mesons. The isovector
meson couples differently to protons and neutrons, thus creating the energy
gap necessary for the direct neutron decay.

In the following we consider a self-consistent relativistic model of nuclear
matter in which baryons, $B=n,p,\Sigma ^{-},\Sigma ^{0},\Sigma ^{+},\Lambda $%
, interact via exchange of $\sigma $, $\omega $, and $\rho $ mesons \cite%
{Serot}. The Lagrangian density,%
\begin{eqnarray}
{\cal L} &=&\sum_{B}\bar{B}\left[ \gamma _{\mu }\left( i\partial ^{\mu
}-g_{\omega B}\omega ^{\mu }-\frac{1}{2}g_{\rho B}{\bf b}^{\mu }\cdot {\bf %
\tau }\right) -\left( M_{B}-g_{\sigma B}\sigma \right) \right] B  \nonumber
\\
&&-\frac{1}{4}F_{\mu \nu }F^{\mu \nu }+\frac{1}{2}m_{\omega }^{2}\omega
_{\mu }\omega ^{\mu }-\frac{1}{4}{\bf B}_{\mu \nu }{\bf B}^{\mu \nu }+\frac{1%
}{2}m_{\rho }^{2}{\bf b}_{\mu }{\bf b}^{\mu }  \nonumber \\
&&+\frac{1}{2}\left( \partial _{\mu }\sigma \partial ^{\mu }\sigma
-m_{\sigma }^{2}\sigma ^{2}\right) -U\left( \sigma \right) +\bar{l}\left(
i\gamma _{\mu }\partial ^{\mu }-m_{l}\right) l,
\end{eqnarray}%
includes the interaction of baryon fields $B$ with a scalar field $\sigma $,
a vector field $\omega _{\mu }$ and an isovector field ${\bf b}_{\mu }$ of $%
\rho $-meson. In the above, $B$ are the Dirac spinor fields for baryons, $%
{\bf b}_{\mu }$ is the isovector field of $\rho $-meson. We denote as ${\bf %
\tau }$ the isospin operator, which acts on the baryons of the bare mass $%
M_{B}$. The leptons are represented only by electrons and muons, $%
l=e^{-},\mu ^{-}$, which are included in the model as noninteracting
particles. The field strength tensors for the $\omega $ and $\rho $ mesons
are $F_{\mu \nu }=\partial _{\mu }\omega _{\nu }-\partial _{\nu }\omega
_{\mu }$ and ${\bf B}_{\mu \nu }=\partial _{\mu }{\bf b}_{\nu }-\partial
_{\nu }{\bf b}_{\mu }$, respectively. The potential $U\left( \sigma \right) $
represents the self-interactions of the scalar field and is taken to be of
the form 
\begin{equation}
U\left( \sigma \right) =\frac{1}{3}bM\left( g_{\sigma N}\sigma \right) ^{3}+%
\frac{1}{4}c\left( g_{\sigma N}\sigma \right) ^{4}.
\end{equation}%
The parameters of the model are chosen as suggested in Ref. \cite{GM} to
reproduce the nuclear matter equilibrium density, the binding energy per
nucleon, the symmetry energy, the compression modulus, and the nucleon
effective mass at saturation density $n_{0}=0.16$ $fm^{-3}$.

In the mean field approximation, when the contribution of mesons reduce to
classical condensate fields $\left\langle \sigma \right\rangle =\sigma _{0}$%
, $\left\langle \omega ^{\mu }\right\rangle =\omega _{0}\delta ^{\mu 0}$, $%
\left\langle {\bf b}^{\mu }\right\rangle \equiv \left( 0,0,\rho _{0}\right)
\delta ^{\mu 0}$, only the baryon fields must be quantized. This procedure
yields the following linear Dirac equation for the nucleon 
\begin{equation}
\left( i\partial _{\mu }\gamma ^{\mu }-g_{\omega }\gamma ^{0}\omega _{0}-%
\frac{1}{2}g_{\rho }\gamma ^{0}\rho _{0}\tau _{3}-\left( M-g_{\sigma }\sigma
_{0}\right) \right) \Psi \left( x\right) =0,  \label{DiracEq}
\end{equation}%
The stationary and uniform condensate fields equally shift the effective
masses 
\begin{equation}
M^{\ast }=M-g_{\sigma }\sigma _{0}  \label{Mef}
\end{equation}%
but lead to different potential energies of protons and neutrons 
\begin{equation}
U_{{\rm n}}=g_{\omega }\omega _{0}-\frac{1}{2}g_{\rho }\rho _{0},\,\ \ \ U_{%
{\rm p}}=g_{\omega }\omega _{0}+\frac{1}{2}g_{\rho }\rho _{0},  \label{Unp}
\end{equation}%
thus creating the energy gap $U_{{\rm n}}-U_{{\rm p}}=-g_{\rho }\rho _{0}$
between possible energies of protons and neutrons.

In the lowest order in the Fermi weak coupling constant $G_{F}$, the matrix
element of the neutron beta decay is found to be%
\begin{eqnarray}
{\cal M}_{fi} &=&-i\frac{G_{F}C}{\sqrt{2}}\bar{u}_{l}\left( k_{2}\right)
\gamma _{\mu }\left( 1+\gamma _{5}\right) \nu \left( -k_{1}\right) \,\times
\label{mend} \\
&&\times \bar{u}_{{\rm p}}\left( P^{\prime }\right) \left[ C_{V}\gamma ^{\mu
}+\frac{1}{2M}C_{M}\sigma ^{\mu \nu }q_{\nu }+C_{A}\left( \gamma ^{\mu
}\gamma _{5}+F_{q}\,q^{\mu }\gamma _{5}\right) \right] u_{{\rm n}}\left(
P\right) ,  \nonumber
\end{eqnarray}%
were $C=\cos \theta _{C}=0\allowbreak .\,\allowbreak 973$ is the Cabibbo
factor. This matrix element includes also the terms caused by weak magnetism
and pseudoscalar interaction. In the mean field approximation, we assume $%
C_{V}=1$, $C_{M}=\lambda _{p}-\lambda _{n}\simeq 3.7$, $C_{A}=1.26$, and%
\begin{equation}
F_{q}\,=-\,\frac{2M^{\ast }}{\left( m_{\pi }^{2}-q^{2}\right) }.  \label{Fq}
\end{equation}%
In the above $\lambda _{p}$ and $\lambda _{n}$ are the anomalous magnetic
moments of the proton and the neutron respectively.

Note that the matrix element (\ref{mend}) is of the same form as that for
the neutron decay in a free space, but with the total momentum transfer
replaced with the kinetic momentum transfer $q=\left( \varepsilon
-\varepsilon ,{\bf p-p}\right) $.

We consider the total energy which is emitted into neutrino and antineutrino
per unit volume and time. Within beta equilibrium, the inverse reaction $%
p+l\rightarrow n+\nu _{l}$ corresponding to a capture of the lepton $l$,
gives the same emissivity as the beta decay, but in neutrinos. Thus, the
total energy loss $Q$ for the Urca processes is twice more than that caused
by the beta decay. Taking this into account by Fermi's ''golden'' rule we
find\footnote{%
Here we rectify an error made in the journal version of the paper. The
correct expression can be obtained from that published in \cite{L2002}, \cite%
{NPA02} by simple replacement $C_{M}\rightarrow C_{M}/2$. The author is
grateful to M. Prakash and S. Ratkovi\'{c} who have pointed out this error.}:%
\begin{eqnarray}
Q &=&\,\frac{457\pi }{10\,080}G_{F}^{2}C^{2}T^{6}\Theta \left( p_{l}+p_{{\rm %
p}}-p_{{\rm n}}\right) \left\{ \left( C_{A}^{2}-C_{V}^{2}\right) M^{\ast
2}\mu _{l}\right.  \nonumber \\
&&+\frac{1}{2}\left( C_{V}^{2}+C_{A}^{2}\right) \left[ 4\varepsilon _{{\rm n}%
}\varepsilon _{{\rm p}}\mu _{l}-\left( \varepsilon _{{\rm n}}-\varepsilon _{%
{\rm p}}\right) \left( \left( \varepsilon _{{\rm n}}+\varepsilon _{{\rm p}%
}\right) ^{2}-p_{l}^{2}\right) \right]  \nonumber \\
&&+C_{V}C_{M}\frac{M^{\ast }}{2M}\left[ 2\left( \varepsilon _{{\rm n}%
}-\varepsilon _{{\rm p}}\right) p_{l}^{2}-\left( 3\left( \varepsilon _{{\rm n%
}}-\varepsilon _{{\rm p}}\right) ^{2}-p_{l}^{2}\right) \mu _{l}\right] 
\nonumber \\
&&+C_{A}\left( C_{V}+\frac{M^{\ast }}{M}C_{M}\right) \left( \varepsilon _{%
{\rm n}}+\varepsilon _{{\rm p}}\right) \left( p_{l}^{2}-\left( \varepsilon _{%
{\rm n}}-\varepsilon _{{\rm p}}\right) ^{2}\right)  \nonumber \\
&&+C_{M}^{2}\frac{1}{16M^{2}}\left[ 8M^{\ast 2}\left( \varepsilon _{{\rm n}%
}-\varepsilon _{{\rm p}}\right) \left( p_{l}^{2}-\left( \varepsilon _{{\rm n}%
}-\varepsilon _{{\rm p}}\right) \mu _{l}\right) \right.  \nonumber \\
&&+\left( p_{l}^{2}-\left( \varepsilon _{{\rm n}}-\varepsilon _{{\rm p}%
}\right) ^{2}\right) \left( 2\varepsilon _{{\rm n}}^{2}+2\varepsilon _{{\rm p%
}}^{2}-p_{l}^{2}\right) \mu _{l}  \nonumber \\
&&\left. -\left( p_{l}^{2}-\left( \varepsilon _{{\rm n}}-\varepsilon _{{\rm p%
}}\right) ^{2}\right) \left( \varepsilon _{{\rm n}}+\varepsilon _{{\rm p}%
}\right) ^{2}\left( 2\varepsilon _{{\rm n}}-2\varepsilon _{{\rm p}}-\mu
_{l}\right) \right]  \nonumber \\
&&\left. -C_{A}^{2}M^{\ast 2}\Phi \left( 1+m_{\pi }^{2}\Phi \right) \left[
\mu _{l}\left( \left( \varepsilon _{{\rm n}}-\varepsilon _{{\rm p}}\right)
^{2}+p_{l}^{2}\right) -2\left( \varepsilon _{{\rm n}}-\varepsilon _{{\rm p}%
}\right) p_{l}^{2}\right] \right\}  \label{QMFA}
\end{eqnarray}%
with $\Theta \left( x\right) =1$ if $x\geq 0$ and zero otherwise. In the
above, $\varepsilon _{{\rm n}}$ and $\varepsilon _{{\rm p}}$ are the kinetic
Fermi energies of neutrons and protons respectively, and the last term, with 
$\allowbreak $ 
\begin{equation}
\Phi =\frac{1}{m_{\pi }^{2}+p_{l}^{2}-\left( \varepsilon _{{\rm n}%
}-\varepsilon _{{\rm p}}\right) ^{2}},  \label{FI}
\end{equation}%
represents the contribution of the pseudoscalar interaction.

Neutrino energy losses caused by the direct Urca on nucleons depend
essentially on the composition of beta-stable nuclear matter. In order to
examine numerically the relativistic effects in the direct Urca processes,
we consider first a simplified model for degenerate nuclear matter of the
standard composition consisting on neutrons, protons, electrons, and muons
under beta-equilibrium. The left panel of Fig. 3 shows the composition of
neutrino-free matter in beta equilibrium among nucleons, electrons and
muons. The fractions of the constituents are shown versus the baryon number
density, in units of saturation density. The right panel represents the
emissivity of the Urca process, as given by our formula in comparison with
the emissivity calculated without the effects of weak magnetism and pseudo
scalar interaction. The short-dashed curve is the non-relativistic
emissivity. All the emissivities are given in units $10^{27}T_{9}^{6}$ $%
erg\,cm^{-3}s^{-1}$, where the temperature $T_{9}=T/10^{9}\,K$. Thus the
curves demonstrate how the emissivity varies along with the matter density.

We can observe that the relativistic effects dramatically modify the
emissivity. The non-relativistic approximation approaches our result only at
densities much smaller than the threshold density, indicated by the vertical
line. Due to the decrease of the nucleon effective mass, the
non-relativistic formula predicts a decreasing of the emissivity as density
increases above the threshold. In contrast, our formula shows a substantial
increasing of the neutrino energy losses as we go to larger densities. The
long-dashed curve demonstrates the energy losses obtained from the
relativistic formula by formal setting $C_{M}$ equal zero and $\Phi $ equal
zero (See also Ref. \cite{PLB}). This eliminates the weak magnetism and
pseudo scalar contributions. A comparison of the long-dashed and solid curve
demonstrates that due to weak magnetism effects, relativistic emissivities
increase by approximately 40-50\%.

\vskip0.3cm \psfig{file=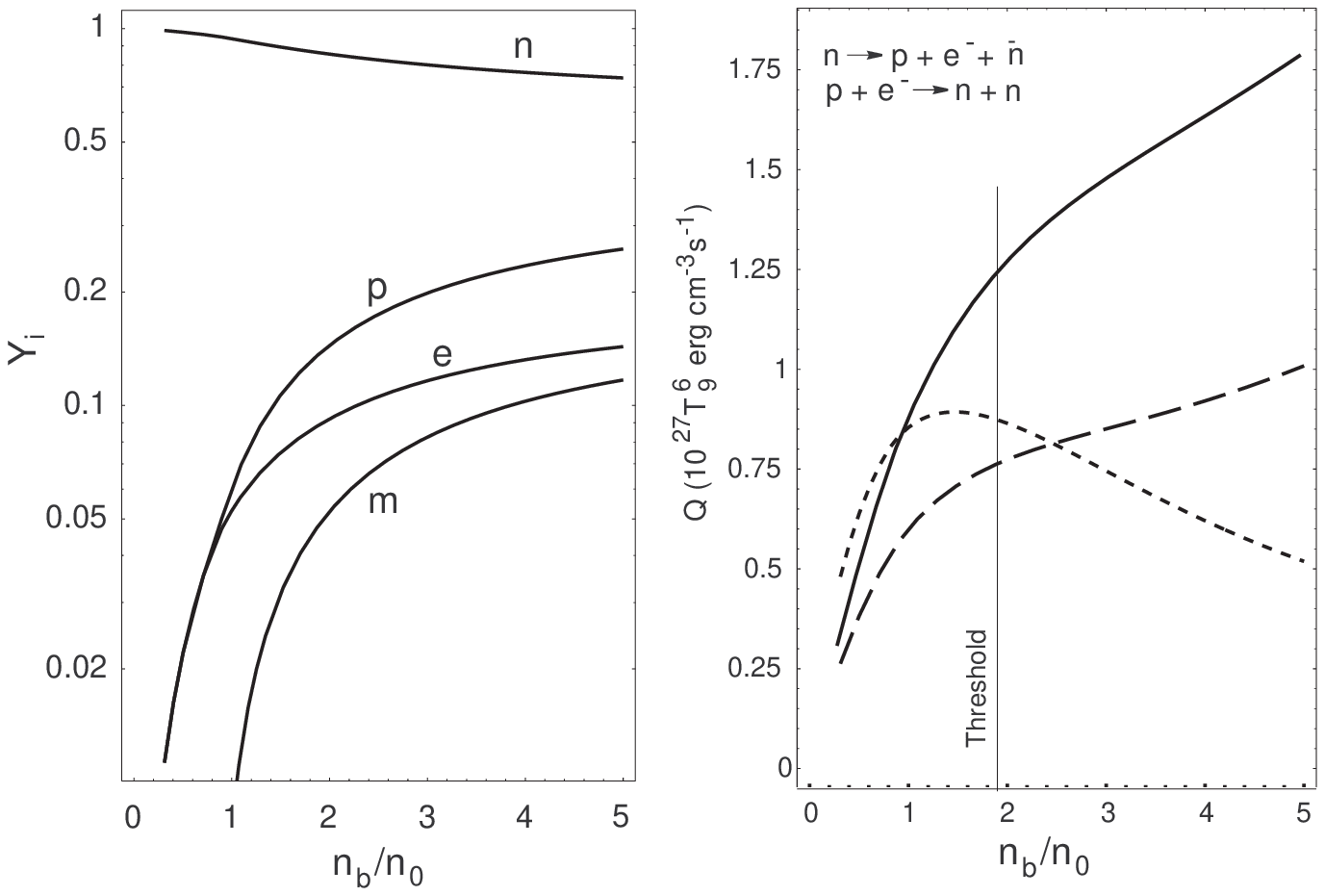}Fig. 3. The left panel shows the
composition of neutrino-free matter in beta equilibrium among nucleons,
electrons and muons. The fractions are shown versus the baryon number
density, in units of saturation density. Right panel represents the
emissivity of the Urca process, as given by our formula (solid curve) in
comparison with the emissivity calculated without the effects of weak
magnetism and pseudo scalar interaction (long-dased curve). The short-dashed
curve is the non-relativistic emissivity. All the emissivities are given in
units $10^{27}T_{9}^{6}$ $erg\,cm^{-3}s^{-1}$, where the temperature $%
T_{9}=T/10^{9}\,K$. \vskip0.3cm

To examine dependence of energy losses on the matter composition we consider
the model, which includes nucleon and hyperon degrees of freedom. The
composition of neutrino-free matter in beta equilibrium among nucleons,
hyperons, electrons and muons is shown in this figure versus the baryon
number density. The right panel represents the emissivity of the direct Urca
processes, as given by our formula in comparison with the emissivity
calculated without the effects of weak magnetism and pseudo scalar. The
short deshed curve is again the non-relativistic emissivity. We see that the
neutrino energy losses caused by the direct Urca processes on nucleons
depend essentially on the composition of beta-stable nuclear matter.
Appearance of hyperons in the system suppresses the nucleon fractions and
lepton abundance. Therefore at densities, where the number of hyperons is
comparable with the number of protons, the relativistic emissivity reaches
the maximum and then has a tendency to decrease. The relativistic
emissivity, however, is found to be substantially larger than that predicted
in the non-relativistic approach. The weak magnetism effects again
approximately doubles the relativistic emissivity.

\vskip0.3cm \psfig{file=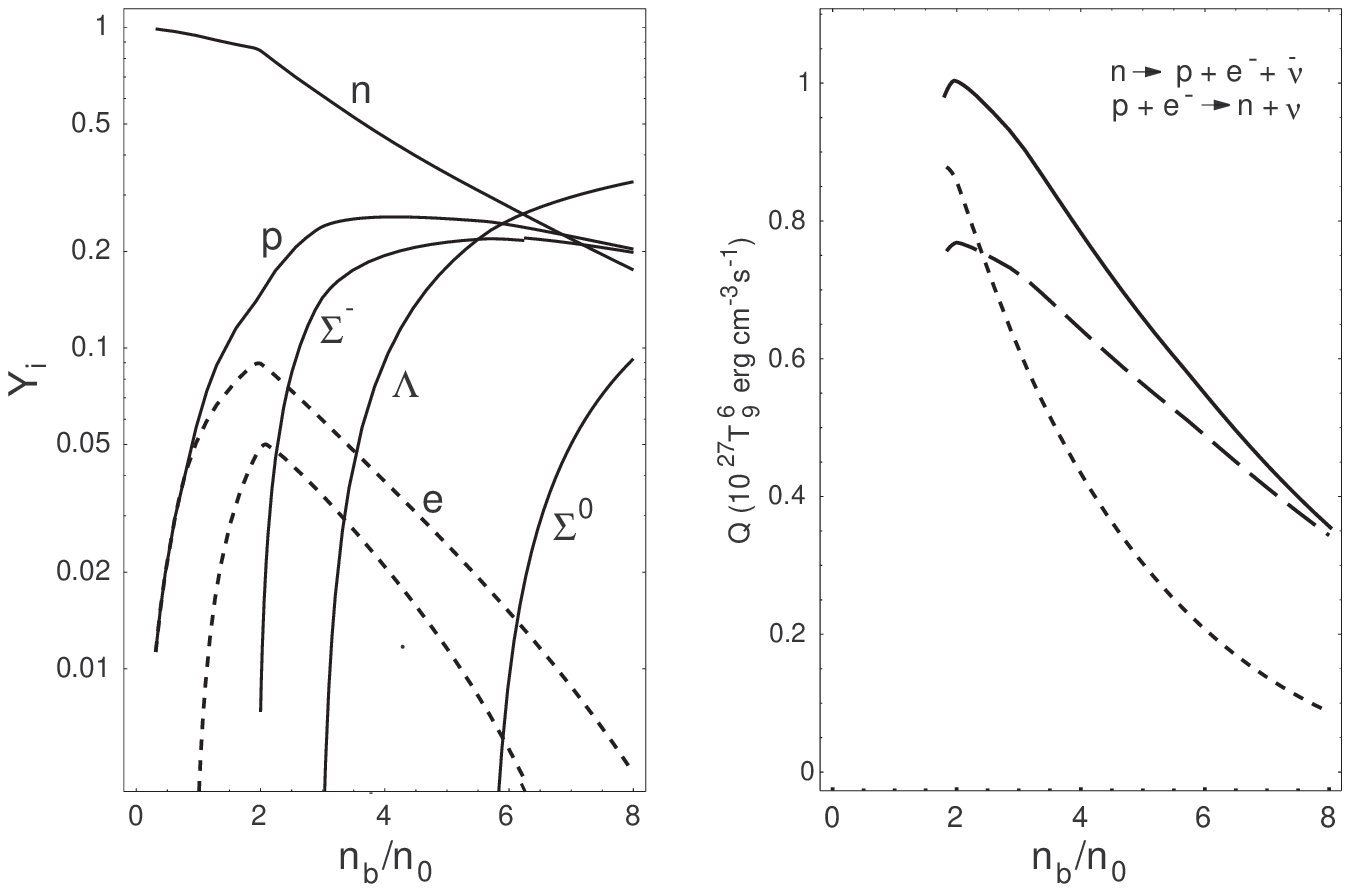}Fig. 4. The same as Fig. 3 but for the
matter composition including hyperons. \vskip0.3cm

Let's summarize our results.We have shown that the direct Urca processes in
a superdense matter of neutron star cores are kinematically allowed only due
to isovector mesons, which differently interact with protons and neutrons.
By creating the energy gap between proton and neutron spectrums the
isovector mesons support a time-like total momentum transfer from the
nucleon, as required by kinematics of the reaction. In the mean field
approximation, we derived the matrix element of the nucleon transition
current, which is found to be a function of the space-like kinetic momentum
transfer. In the mean field approximation, we have calculated the neutrino
energy losses caused by the direct Urca processes on nucleons. Our Eq. (\ref%
{QMFA}) for neutrino energy losses exactly incorporates the effects of
nucleon recoil, parity violation, weak magnetism, and pseudoscalar
interaction. To quantify the relativistic effects we consider a
self-consistent relativistic model, widely used in the theory of
relativistic nuclear matter. The relativistic energy losses are up to four
times larger than those given by the non-relativistic approach. In our
analysis, we pay special attention to the effects of weak magnetism and
pseudoscalar interaction in the neutrino energy losses. We found that due to
weak magnetism effects, relativistic emissivities increase by approximately
40-50\%. The efficiency of the direct Urca processes involving different
kinds of baryons depends essentially on the composition of the beta-stable
nuclear matter. The energy losses in the standard nuclear matter increase
along with the density. In contrast, the total energy losses in the
multicomponent medium have a tendency to decrease.


\begin{references}
\bibitem{Lat91} J. M. Lattimer, C.J. Pethick, M. Prakash, and P. Haensel,
Phys. Rev. Lett., 66 (1991) 2701.

\bibitem{Serot} B. D. Serot and J. D. Walecka, Adv. Nucl. Phys. 19, eds.
J.W. Negele and E. Vogt, (Plenum, New York, 1986); B.D. Serot, Rep. Prog.
Phys. 55 (1992) 1855.

\bibitem{GM} N. K.Glendenning and S. A. Moszkowski, Phys. Rev. Lett. 67
(1991) 2414.

\bibitem{L2002} L. B. Leinson, Phys. Lett. B 532 (2002) 267;

\bibitem{NPA02} L. B. Leinson, Nucl, Phys. A 707 (2002) 543.

\bibitem{PLB} L. B. Leinson and A. Perez, Phys. Lett. B 518 (2001) 15;
Erratum: Phys. Lett. B, 522 (2001) 358.
\end{references}
\end{document}